\documentclass{PoS}

\title{Partial deconfinement in gauge theories
\\ \vspace{-60mm}\hspace{13cm} \small{\textrm{UTHEP-739}} \vspace{60mm}
}

\ShortTitle{Partial deconfinement in gauge theories}

\author{Masanori Hanada\\
        STAG Research Centre, University of Southampton, Southampton, SO17 1BJ, U.K.\\
        E-mail: \email{m.hanada@soton.ac.uk}}

\author{Goro Ishiki\\
        Tomonaga Center for the History of the Universe, University of Tsukuba, and\\
        Graduate School of Pure and Applied Sciences University of Tsukuba, Tsukuba, Ibaraki 305-8571, Japan\\
        E-mail: \email{Ishiki@het.ph.tsukuba.ac.jp}}

\author{\speaker{Hiromasa Watanabe}\\
        Graduate School of Pure and Applied Sciences, University of Tsukuba, Tsukuba, Ibaraki 305-8571, Japan\\
        E-mail: \email{watanabe@het.ph.tsukuba.ac.jp}}

\abstract{We provide the evidence for the existence of {\it partially deconfined phase} in large-$N$ gauge theory. 
In this phase, the SU($M$) subgroup of SU($N$) gauge group deconfines, where $\frac{M}{N}$
changes continuously from zero (confined phase) to one (deconfined phase). 
The partially deconfined phase may exist in real QCD with $N=3$.}

\FullConference{37th International Symposium on Lattice Field Theory - Lattice2019\\
		16-22 June 2019\\
		Wuhan, China}

\begin{document}
\section{Introduction}
\label{sec:intro}
The mechanism of deconfinement transition in QCD is one of the profound problems extensively studied among lattice community. 
In this talk, we consider the large-$N$ gauge theory, which is often a good approximation to finite $N$, including $N=3$. 
We provide the evidence for the existence of a new phase between confined and deconfined phases, which we call ``partially deconfined phase'', 
and discuss the implication to real QCD with $N=3$.     

Let us start the discussion with the most basic feature of the large-$N$ gauge theory: the 't Hooft counting \cite{tHooft:1973alw}. 
In the 't Hooft large-$N$ limit (i.e.~$\lambda=g_{\rm YM}^2N$ and the energy scales under consideration are fixed), 
the energy $E$ and entropy $S$ scales as $N^2$ in the deconfined phase, while they are of order $N^0$ in the confined phase.\footnote{
Strictly speaking, this counting holds when all the fields are in the adjoint representation. 
With fundamental quarks, the counting changes slightly, but the argument does not change essentially. 
} 
Physically, it simply counts the number of degrees of freedom; namely, in order to excite $N^2$ color degrees of freedom, 
the energy of order $N^2$ is needed. 

This simple argument raises a question: what happens if the energy is somewhere between $N^0$ and $N^2$, 
say $E\sim\frac{N^2}{100}$ or $\frac{N^2}{10000000000}$? 
Because the energy is not small enough, the system cannot be in the confined phase; 
at the same time, because the energy is not large enough, it cannot be in the deconfined phase; 
hence something in between has to be realized. 
We propose that {\it partial deconfinement} takes place then. 
Namely, SU$(\frac{N}{10})$ or SU$(\frac{N}{100000})$ subgroup of SU($N$) deconfines \cite{Hanada:2016pwv,Hanada:2018zxn,Berenstein:2018lrm,Hanada:2019czd}. 
More generally, when $E\sim \epsilon N^2$, we expect that SU($M$)-subgroup, with $M\sim\sqrt{\epsilon}N$, deconfines. 
For a heuristic argument that partial deconfinement is a plausible scenario, see \cite{Hanada:2019kue}. 

Partial deconfinement is not as exotic as it may appear at the first glance. Simply, confined and deconfined phases are coexisting 
in the space of color degrees of freedom, or {\it internal space}. 
Similar coexistence of two phases is very common in usual space, which we denote as {\it physical space}. 
For example, at 1-atm and zero-celsius, liquid water and ice coexist.  
It is nothing special, we can see it in a freezer; then why don't we expect the same in the internal space?

In the internal space, the interaction is nonlocal, in that all components of the fields interact with each other via the nonlinear interaction. 
On the other hand, in the physical space, interactions are usually sufficiently local. 
This causes big differences, as we will see. 
\section{Explicit demonstration for weakly-coupled 4D SU($N$) Yang-Mills theories on S$^3$}
\label{sec:4dYMonS3}
\begin{figure}[t]
 \begin{minipage}{0.32\hsize}
 \begin{center}
  \includegraphics[width=37mm,pagebox=cropbox]{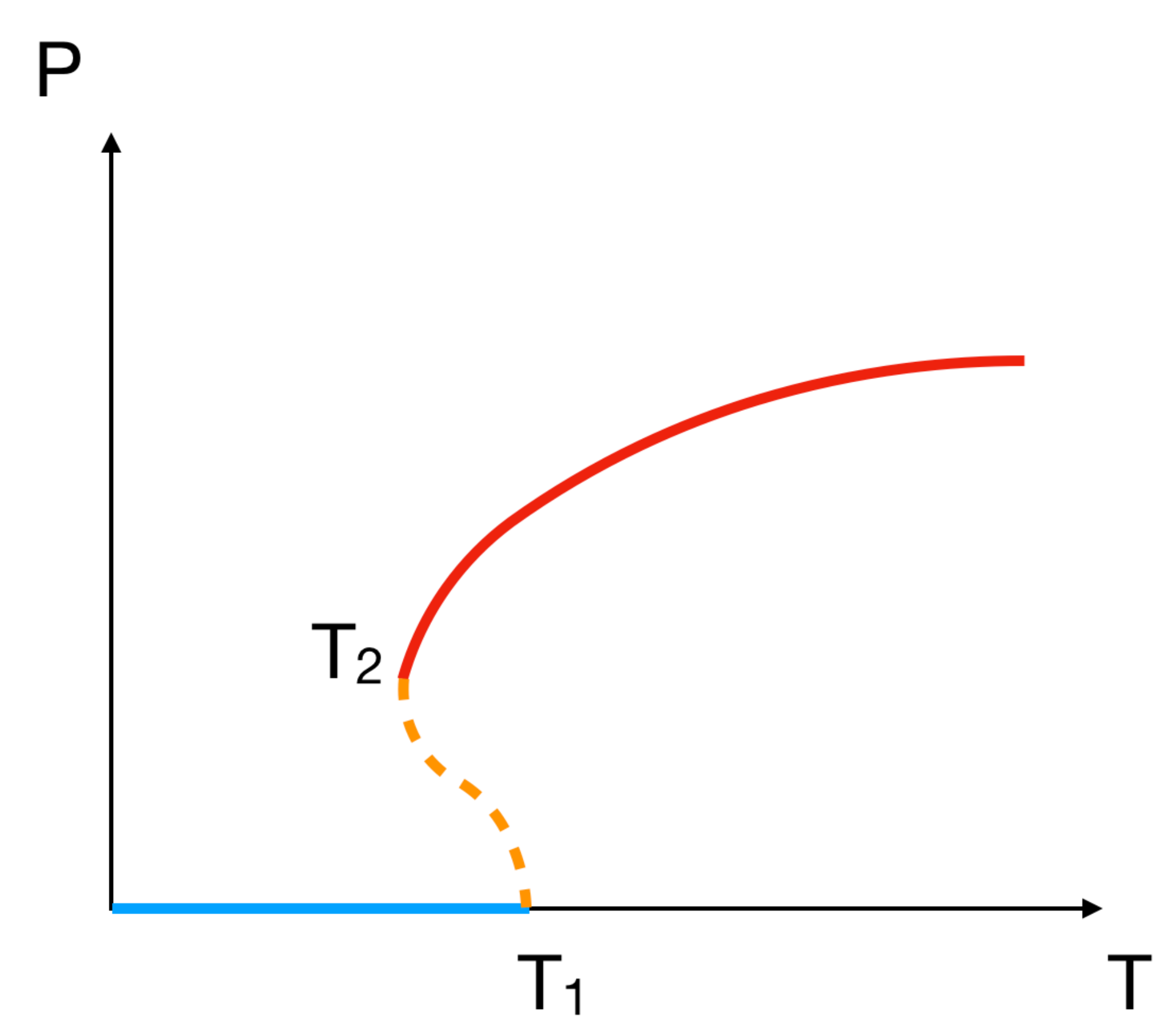}
 \end{center}
 \end{minipage}
 \begin{minipage}{0.32\hsize}
 \begin{center}
  \includegraphics[width=37mm,pagebox=cropbox]{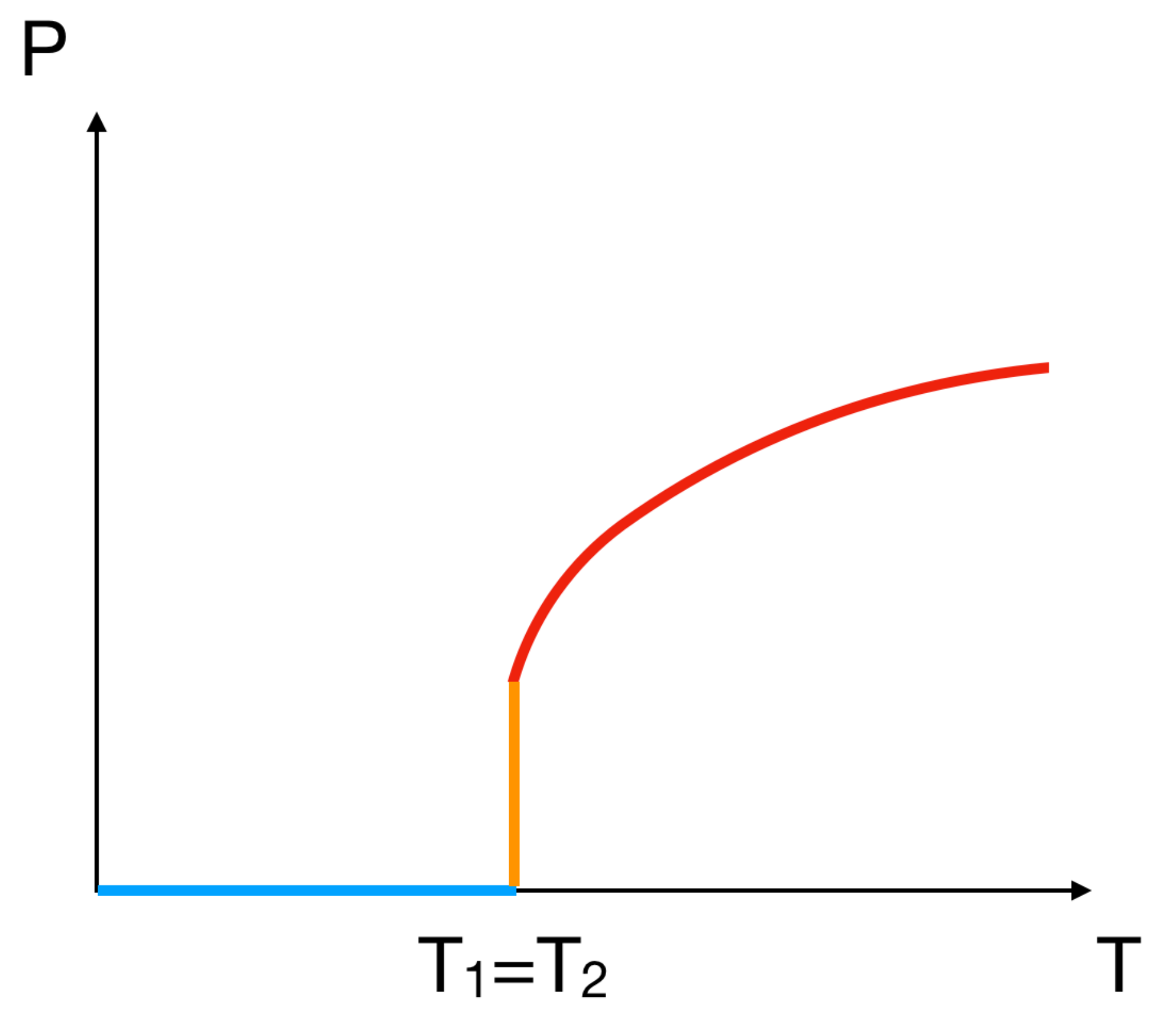}
 \end{center}
 \end{minipage}
  \begin{minipage}{0.32\hsize}
 \begin{center}
  \includegraphics[width=37mm,pagebox=cropbox]{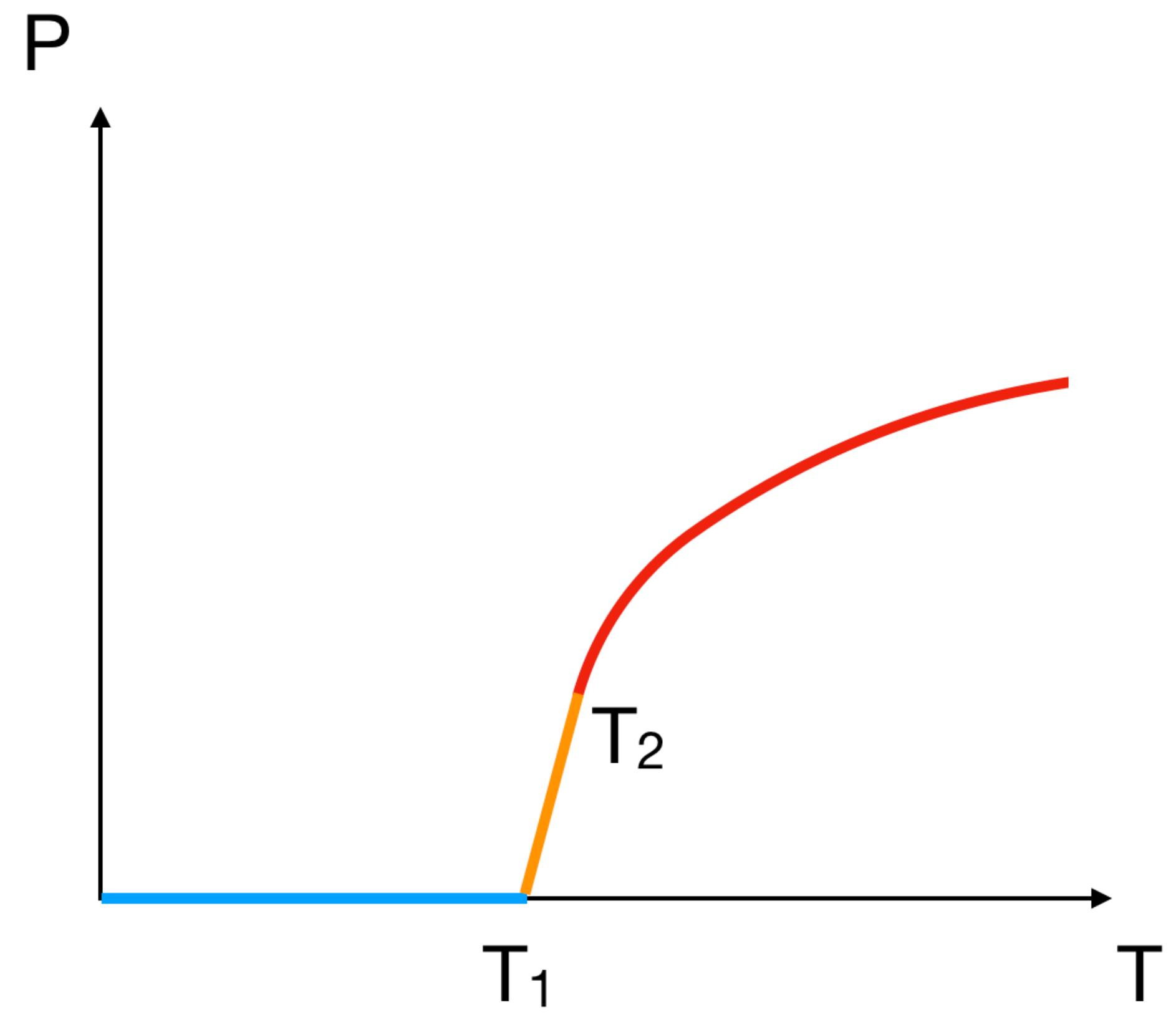}
 \end{center}
 \end{minipage}
 \caption{
Cartoon pictures of three basic types of the phase structure, with the Polyakov loop $P$ as a function of temperature $T$. 
 The blue, orange and red lines represent the confined, partially deconfined and completely deconfined phases.  
 Similar graphs can be drawn even if changing the y-axis to the energy or entropy as well. 
 }
     \label{fig:PvsT}
\end{figure}

Let us consider weakly-coupled 4D SU($N$) Yang-Mills theories on S$^3$ \cite{Sundborg:1999ue,Aharony:2003sx,Hanada:2019czd} as a concrete and solvable example. 
In the weak-coupling limit, the partition function of the whole theory can be described by the unitary matrix model \cite{Sundborg:1999ue,Aharony:2003sx}.  
The extension adding the contribution from finite coupling has been also argued \cite{Aharony:2005bq}. 
Despite its simplicity, this setup has rich properties resembling the large-volume, strongly-coupled theory.
In the weak-coupling limit, the Polyakov loop behaves like the center panel of Fig. \ref{fig:PvsT}. 
The vertical orange line is naturally identified with the partially deconfined phase. 
Very importantly, {\it there are two transitions}: 
one from confinement to partial deconfinement ($P=0$), and the other from partial deconfinement to complete deconfinement ($P=\frac{1}{2}$). 
These transitions are the Hagedorn transition \cite{Hagedorn:1965st} and Gross-Witten-Wadia (GWW) transition \cite{Gross:1980he,Wadia:2012fr}, respectively \cite{Sundborg:1999ue,Aharony:2003sx}. 

\begin{figure}[t]
\begin{minipage}{0.48\hsize}
 \begin{center}
  \includegraphics[width=50mm,pagebox=cropbox,angle=-90]{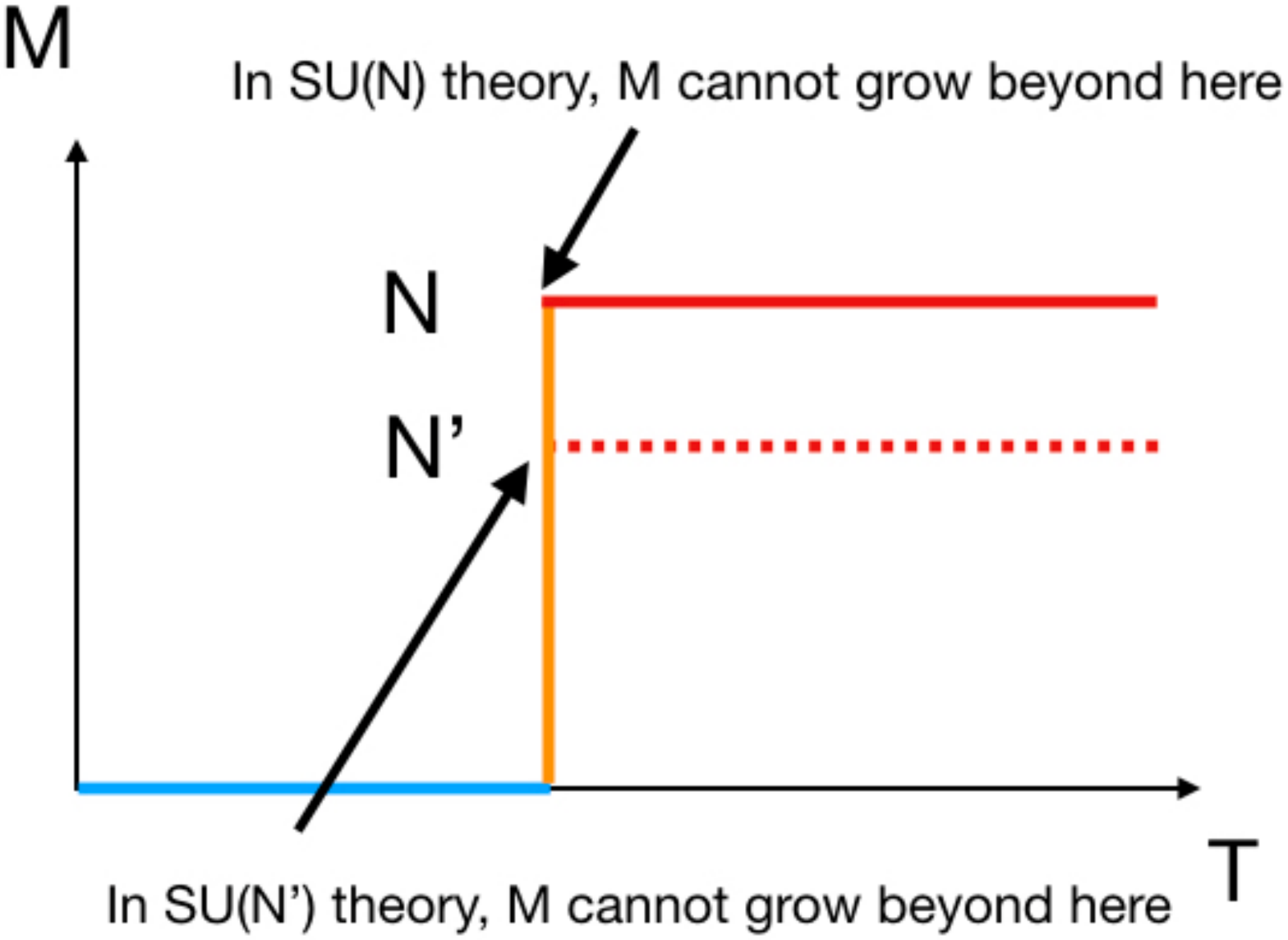}
   \end{center}
 \end{minipage}
 \begin{minipage}{0.48\hsize}
 	\begin{center}
		\includegraphics[width=60mm,pagebox=cropbox]{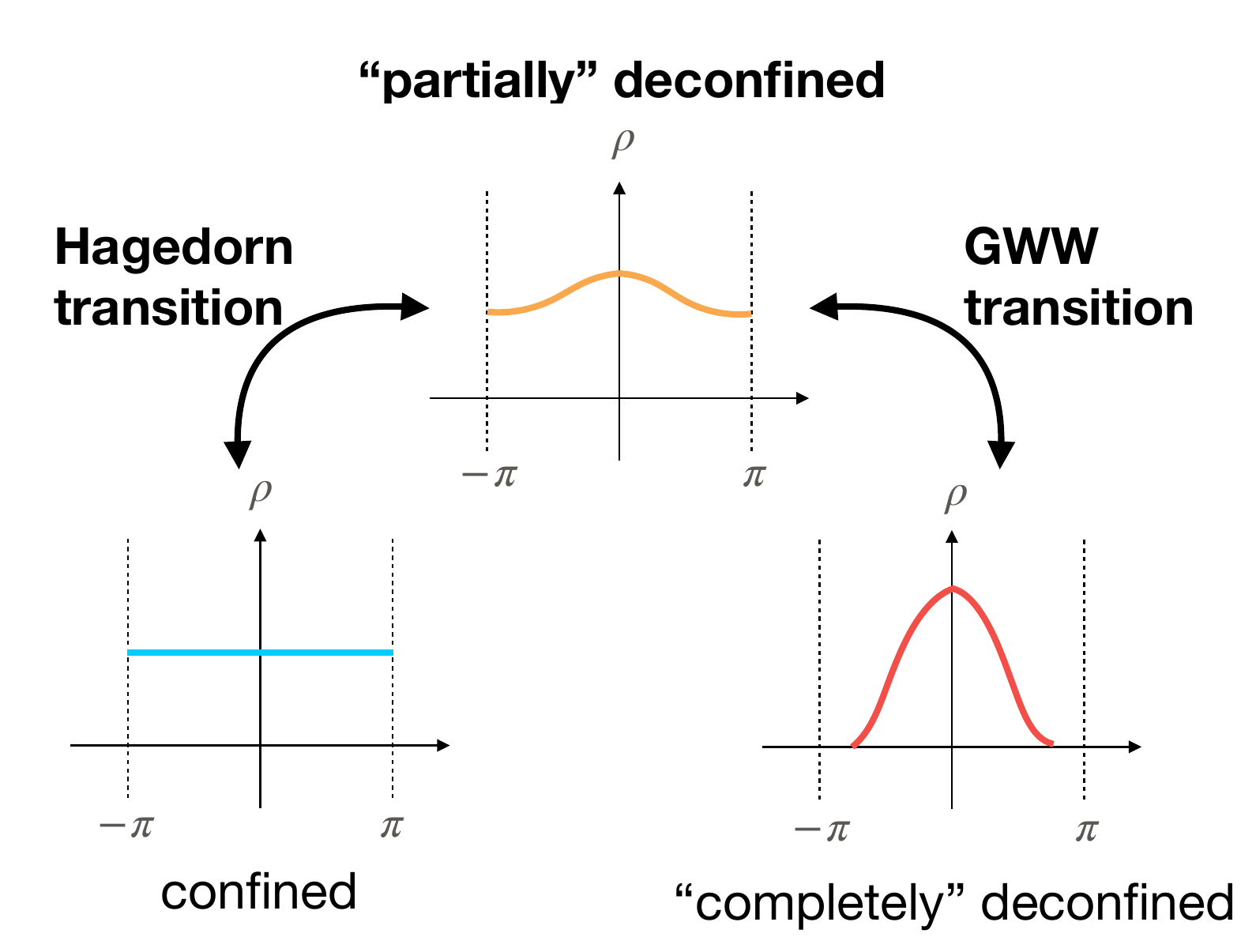}
	\end{center}
\end{minipage}
   \caption{
 [Left] The deconfined sector of SU($N$) theory and SU($N'$) behave in the same way up to $M=N'$. 
 From this, it follows that the SU($M$)-deconfined sector in the SU($N$)-theory corresponds to the GWW-transition point of the SU($M$) theory. 
 [Right] The phase distribution of Polyakov loop in three phases.
  The distribution in partially deconfined phase can be expressed as a sum of the distributions in confined phase and at the GWW-transition point with an appropriate weight.
  The GWW transition corresponds to the formation of a gap at $\theta=\pm\pi$. 
 }
      \label{fig:M-vs-T-free-YM}
\end{figure}

We can find nontrivial evidence of this picture as follows. 
Suppose we had two theories with different gauge group SU($N$) and SU($N'$), where $N'<N$. 
We assume both of them are sufficiently large
such that $1/N$ and $1/N'$ corrections are negligible.
As the energy is pumped up to the system, the size of the deconfined sector can increase up to $M=N$ and $M=N'$, respectively (the left panel of Fig.~\ref{fig:M-vs-T-free-YM}).  
At $M\le N'$, the deconfined sectors in two theories behave in exactly the same manner; 
in other words, the SU($N'$)-deconfined sector in the SU($N$) theory corresponds to the GWW-transition point of the SU($N'$) theory. 
Because the energy $E$ and entropy $S$ are dominated by the deconfined sector, we obtain simple relations which hold in the SU($M$)-partially-deconfined phase \cite{Hanada:2016pwv,Hanada:2018zxn,Hanada:2019czd}: 
\begin{equation}
	E = E_{\rm GWW}(M),\quad S=S_{\rm GWW}(M). 
	\label{eq:consistency_energy_entropy}
\end{equation}
Another consistency condition can be found by looking at the phase distribution of Polyakov loop. 
Because the SU($M$)-deconfined sector corresponds to the GWW-transition point of the SU($M$) theory, 
$M$ out of $N$ phases should be distributed as the GWW-transition point of the SU($M$) theory; we denote this distribution as $\rho_{\mathrm { GWW },M}(\theta)$. 
The other $N-M$ phases should be distributed as the confinement phase.
In the confinement phase, the phases are distributed uniformly, so that Polyakov loop becomes zero: $\rho_{\mathrm { confine }}(\theta)=\frac{1}{2\pi}$. 

Hence, the partial deconfinement implies the following phase distribution \cite{Hanada:2018zxn}: 
\begin{equation}
        \rho(\theta)=\frac{N-M}{N} \rho_{\mathrm { confine }}(\theta)+\frac{M}{N} \rho_{\mathrm { GWW },M}(\theta)=\frac{N-M}{N} \cdot \frac{1}{2 \pi}+\frac{M}{N} \rho_{\mathrm { GWW },M}(\theta).
        \label{eq:consistency_polyakov_loop}
\end{equation}
Note that (\ref{eq:consistency_energy_entropy}) and (\ref{eq:consistency_polyakov_loop}) should be satisfied by the same $M$.
We can check these consistency conditions by explicit calculations \cite{Hanada:2018zxn,Hanada:2019czd}. 
Here, let us see how (\ref{eq:consistency_polyakov_loop}) can be confirmed. 
The distribution of the phases of Polyakov loop is \cite{Sundborg:1999ue,Aharony:2003sx}
\begin{equation}
    \rho(\theta)=
        \left\{
        \begin{array}{cc}
                \displaystyle\frac{1}{2\pi} & (T\le T_1=T_2)\\
                \displaystyle\frac{1}{2\pi}\left(1+A\cos\theta\right) & (T=T_1=T_2)\\
                \displaystyle\frac{A}{\pi }\cos\frac{\theta}{2}\sqrt{\frac{1}{A}-\sin^2\frac{\theta}{2}} \qquad& (T\ge T_1=T_2, |\theta|<2\arcsin\sqrt{A^{-1}})
        \end{array}
        \right. 
\label{eq:GWW}      
\end{equation} 
The Hagedorn transition \cite{Hagedorn:1965st} and Gross-Witten-Wadia (GWW) transition \cite{Gross:1980he,Wadia:2012fr} take place at $A=0$ and at $A=1$, respectively.  
Between these transitions, the distribution is
\begin{equation}
	\rho_\mathrm{p.d.}(\theta) 
	= \left(1-A\right)\cdot \frac1{2\pi} 
		+ A\cdot\frac1{2\pi}\left(1+\cos\theta \right)
	= \left(1-A\right) \rho_\mathrm{confine}
		+ A\cdot\rho_\mathrm{GWW}(\theta).
	\label{eq:4DadjointYM}
\end{equation}
Therefore, it is natural to assume 
\begin{equation}
	A = \frac{M}{N}.  
\end{equation}
It is easy to check that (\ref{eq:consistency_energy_entropy}) is satisfied by the same identification. 

The argument above is a consistency check. In order to demonstrate the partial deconfinement more explicitly, 
it is possible to construct the states in the Hilbert space which dominate the thermodynamics \cite{Hanada:2019czd}. 
The same argument can be repeated with the fundamental quarks \cite{Hanada:2019kue}. 

\section{Further test in matrix quantum mechanics}
\begin{figure}[t]
    	\begin{center}
    		\includegraphics[width=60mm,pagebox=cropbox]{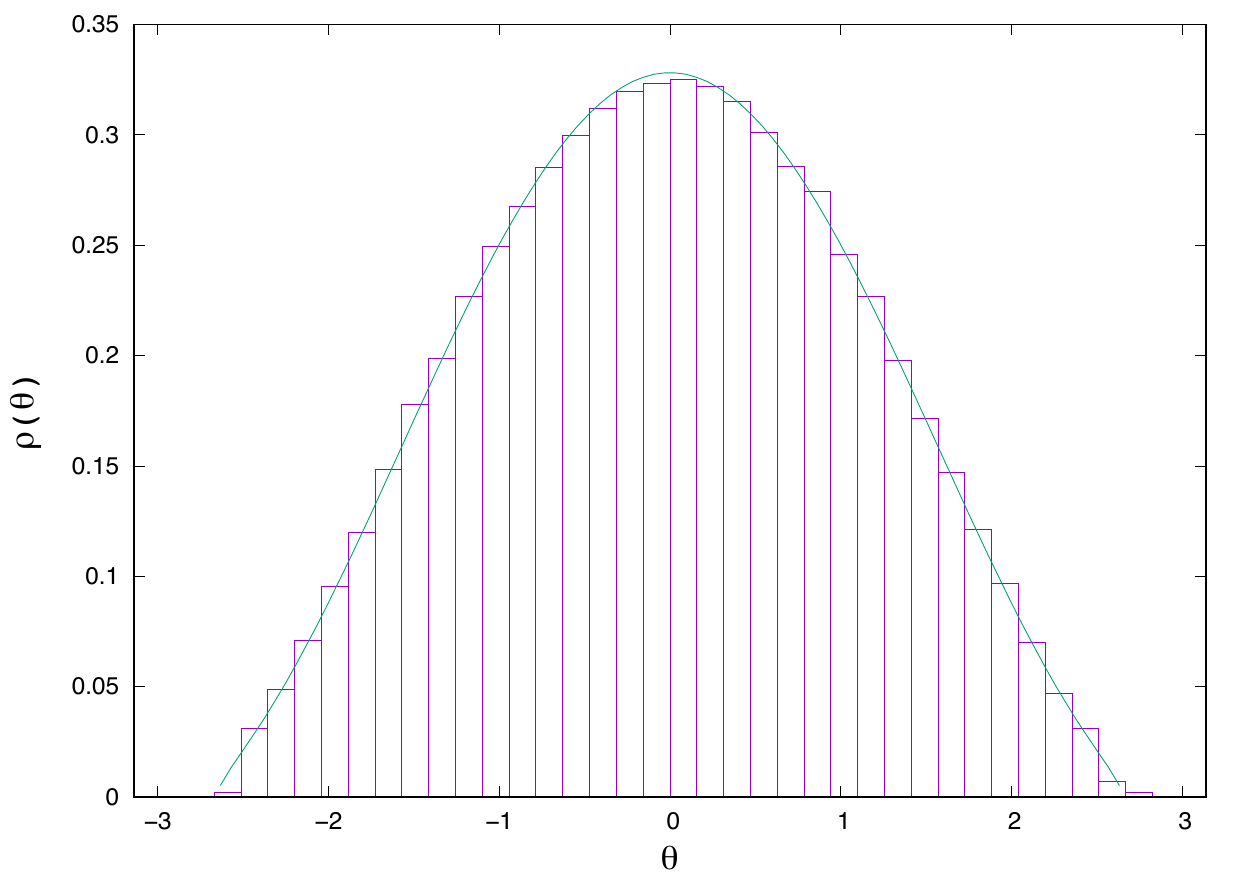}
	\end{center}
    	\caption{The distribution of Polyakov line phase, $N=128$ and $\mu = 5.0$. The number of lattice point is $L=16$. 
	 Temperature is $T = 1.54$, which is slightly above $T_2$. We have used the fit function $\rho(\theta)=\frac{A}{\pi}\cos\frac{\theta}{2}\sqrt{\frac{1}{A}-\sin^2\frac{\theta}{2}}$ and obtained $A \simeq 1.06$.
	}
    \label{fig:BMN_histogram}
\end{figure}

Next let us consider the bosonic part of the plane wave matrix model \cite{Berenstein:2002jq} as a toy model of strongly-coupled gauge theory: 
\begin{eqnarray}
        L = 
        N{\rm Tr}\Biggl(
        \frac{1}{2}\sum_{I=1}^9\left(D_t X_I\right)^2
        &+&\frac{1}{4}\sum_{I,J=1}^9[X_I,X_J]^2 \nonumber\\
        -\frac{\mu^2}{2}\sum_{i=1}^3X_i^2 
        &-&\frac{\mu^2}{8}\sum_{a=4}^9X_a^2 -\mathrm{i}\sum_{i,j,k=1}^3\mu\epsilon^{ijk}X_iX_jX_k\Biggl). 
\end{eqnarray}
Here $X_I$  $(I=1,\cdots,9)$ are $N\times N$ Hermitian matrices and $D_t X_I = \partial_t X_I - \mathrm{i} [A_t, X_I]$. 
This model has the hysteresis like the left panel of Fig.\ref{fig:PvsT}.
At a temperature slightly above $T_2$, we have plotted the histogram of the Polyakov line phases given by our numerical simulation and fitted it by the GWW ansatz (\ref{eq:GWW}). 
As shown in Fig.\ref{fig:BMN_histogram}, the non-uniformly, gapped distribution can be obtained with the fit parameter $A$ is near but larger than 1, which is consistent with partial deconfinement.  
See Refs.~\cite{Hanada:2018zxn,Bergner:2019rca} for more analyses.

\section{Application to QCD}
\label{sec:QCD}
In QCD, neither center symmetry nor chiral symmetry is exact,
due to the presence of the fundamental quarks with nonzero mass. 
Hence, it is widely believed that the deconfinement transition cannot be characterized based on symmetry. 
Partial deconfinement challenges this folklore. 

Let us consider QCD in the large-$N_c$ limit, with $N_f$ fundamental quarks.
At weak coupling and with the compactification on S$^3$, the theory can be solved analytically, 
and the partial deconfinement can be demonstrated explicitly \cite{Hanada:2019kue}.
In the partially deconfined phase, SU($N_c$) gauge symmetry breaks spontaneously
to SU($M_c$)$\times$SU($N_c-M_c$)$\times$U(1) (for the precise meaning of the gauge symmetry breaking, see Ref.~\cite{Hanada:2019czd}). 
Therefore, deconfinement can be characterized by the breaking and restoration of gauge symmetry, associated with the partially broken gauge symmetry in the partially deconfined phase. 

What happens at $N_c=3$? Phase transitions can take place even at finite $N_c$, by sending volume to infinity. 
In order to estimate the amount of the $1/N_c$ corrections, lattice simulation is needed. 
At least, some numerical results \cite{Gupta:2007ax,Mykkanen:2012ri} for 4D SU($N_c$) pure Yang-Mills theory ($N_c=3$ to $6$) suggest
the value of Polyakov loop is close to $\frac12$ at $T=T_2$, which is consistent with the GWW ansatz discussed above. 
(Note however that this ansatz is not a mandatory requirement. Indeed, different phase distribution can also appear 
depending on the detail of the theory \cite{Hanada:2019czd,Hanada:2019kue}.)

For actual QCD at small chemical potential, the thermal `transition' is believed to be a rapid crossover \cite{Aoki:2006we}. 
This region is a good candidate for the partially deconfined phase. 
Recent lattice studies \cite{Denissenya:2014poa,Rohrhofer:2019qwq,Alexandru:2019gdm} which reported peculiar properties in this region may be explained as consequences of partial deconfinement.

\section{Connection to superstring theory}
\label{sec:PD_basics}
\begin{figure}[t]
    \begin{minipage}{0.48\hsize}
        \begin{center}
            \includegraphics[width=60mm, pagebox=cropbox]{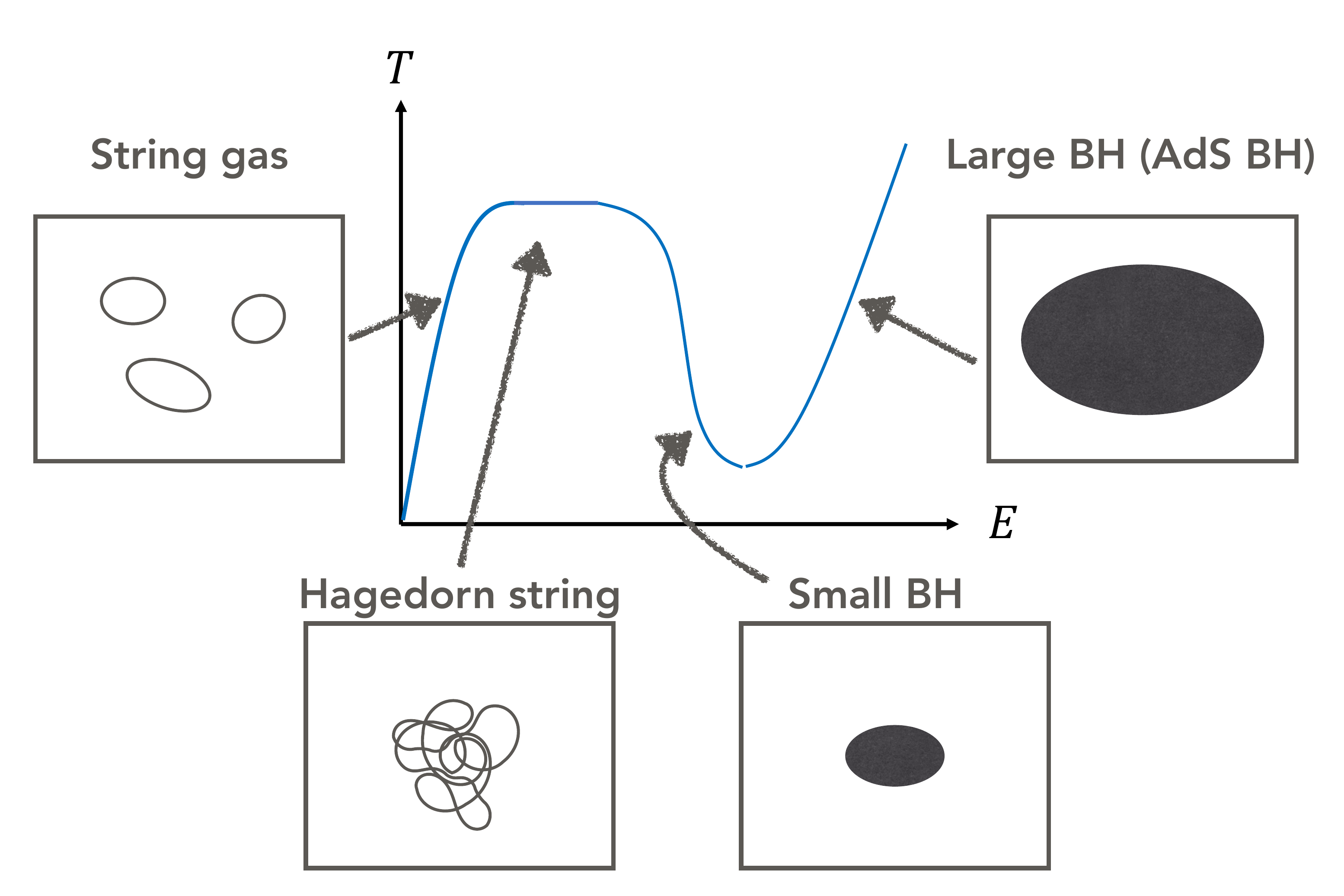}
        \end{center}
    \end{minipage}
    \begin{minipage}{0.48\hsize}
        \begin{center}
            \includegraphics[width=60mm,pagebox=cropbox]{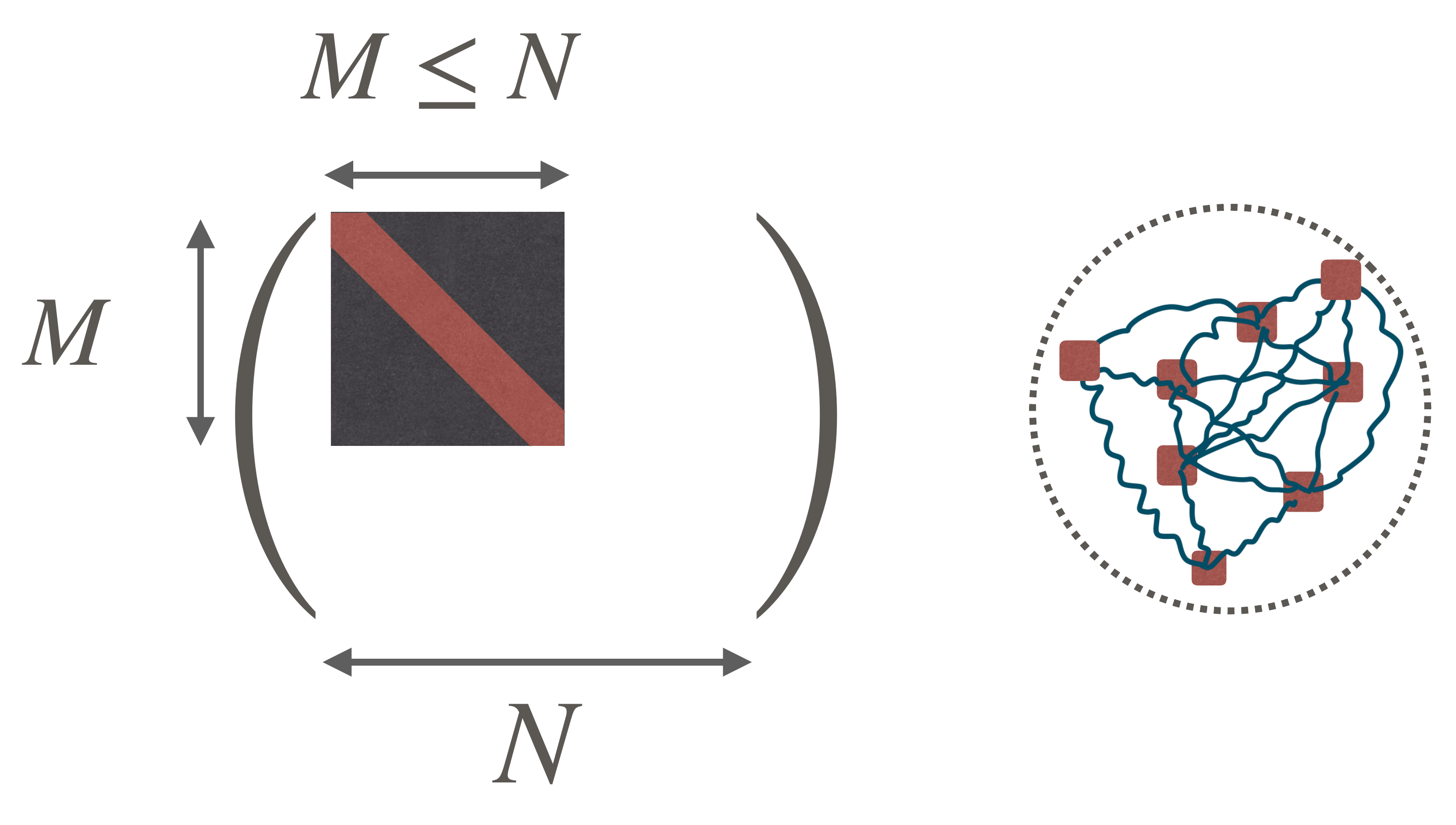}
        \end{center}
    \end{minipage}
        \caption{
        [Left] The phase structure of dual gravity on AdS$_5 \times$S$^5$.
        At high energy, a ``large black hole'' in which is proportional to $T^4$ is formed.
        At low energy, the system can be described by the gas of strings or gravitons. 
        In the middle region, it is known the ``small black hole'' and Hagedorn string is formed.
        The small black hole has negative specific heat since $E \propto N^2T^{-7}$.
       [Right] The  cartoon picture of the scalar fields in the partially deconfined phase. 
       $M$ D-branes are forming a bound state, which is the counterpart of the small black hole.
        }
    \label{fig:phase_diagram}
\end{figure}

Originally, partial deconfinement has been proposed in order to understand a peculiar feature of black hole in the context of string theory \cite{Hanada:2016pwv}.
AdS/CFT duality conjecture \cite{Maldacena:1997re} relates type IIB superstring on AdS$_5 \times$S$^5$ and 4D $\mathcal{N}=4$ SU($N$) super Yang-Mills theory (SYM). 
Thermodynamic property of 4D SYM compactified on S$^3$ can be studied by using various solutions in the gravity side. 
In particular, deconfinement is interpreted as the formation of black hole \cite{Witten:1998zw}. 
The phase structure of the gravity side is shown as Fig.\ref{fig:phase_diagram}. 

If the duality conjecture is correct, the 4D SYM has to explain the same phase structure.  
In particular, the phase corresponding to the small black hole phase, which has negative specific heat ($E\propto N^2T^{-7}$), has to exist . 
But how can such a phase exist in a healthy quantum theory? 

Intuitively, higher temperature means larger energy per degree of freedom. 
If the small black hole is described by the completely deconfined phase, 
the energy per degree of freedom is $\frac{E}{N^2}$. Here $N^2$ is a fixed value, 
and hence, if $E<E'$ then $\frac{E}{N^2}<\frac{E'}{N^2}$ and hence $T<T'$. 
However, if the small black hole is described by the partially deconfined phase, 
the size of the deconfined sector $M$ can change nontrivially as a function of the energy, 
and hence, $\frac{E}{M^2(E)}$ can increase or decrease with $E$ depending on the details of the theory \cite{Berkowitz:2016znt,Hanada:2016pwv}.  
By using this idea, the negative specific heat can be derived, modulo a few technical assumptions \cite{Hanada:2016pwv}.  

%

\section{Summary and Discussion}
We have proposed the existence of the intermediate phase between confined and deconfined phases for generic large-$N$ gauge theory, 
and provided the evidence for a few concrete examples. 
Our proposal explains the actual physical meaning of Gross-Witten-Wadia transition. 
We have argued that the partially deconfined phase, and associated spontaneous breaking of gauge symmetry, may exist in SU($3$) QCD.  
Lattice simulation should be able to test this proposal, and if it is actually happening,  
it would be interesting if we could draw the consequences in collider experiments or cosmological observations. 

Another interesting direction is to combine this idea with the lattice approach to quantum gravity via holographic duality (see e.g.~Refs.\cite{Anagnostopoulos:2007fw,Catterall:2008yz}). 
By properly fixing the gauge configuration-by-configuration, in a way that deconfined and confined sectors are separated,   
we might be able to understand how the geometry emerges from the microscopic degrees of freedom in gauge theory. 

\acknowledgments
The work of M.~H. was partially supported by the STFC Ernest Rutherford Grant ST/R003599/1 and JSPS  KAKENHI  Grants17K1428. 
The work of G.~I. was supported, in part, by Program to Disseminate Tenure Tracking System, MEXT, Japan and by KAKENHI (16K17679 and 19K03818).

\bibliographystyle{JHEP_mod}
\bibliography{ref}

\providecommand{\href}[2]{#2}\begingroup\raggedright\begin{thebibliography}{10}

\bibitem{tHooft:1973alw}
G.~'t~Hooft, \href{https://doi.org/10.1016/0550-3213(74)90154-0}{\emph{Nucl.
  Phys.} {\bfseries B72} (1974) 461}.

\bibitem{Hanada:2016pwv}
M.~Hanada and J.~Maltz,
  \href{https://doi.org/10.1007/JHEP02(2017)012}{\emph{JHEP} {\bfseries 02}
  (2017) 012} [\href{https://arxiv.org/abs/1608.03276}{{\ttfamily
  1608.03276}}].

\bibitem{Hanada:2018zxn}
M.~Hanada, G.~Ishiki and H.~Watanabe,
  \href{https://doi.org/10.1007/JHEP03(2019)145}{\emph{JHEP} {\bfseries 03}
  (2019) 145} [\href{https://arxiv.org/abs/1812.05494}{{\ttfamily
  1812.05494}}].

\bibitem{Berenstein:2018lrm}
D.~Berenstein, \href{https://doi.org/10.1007/JHEP09(2018)054}{\emph{JHEP}
  {\bfseries 09} (2018) 054}
  [\href{https://arxiv.org/abs/1806.05729}{{\ttfamily 1806.05729}}].

\bibitem{Hanada:2019czd}
M.~Hanada, A.~Jevicki, C.~Peng and N.~Wintergerst,
  \href{https://arxiv.org/abs/1909.09118}{{\ttfamily 1909.09118}}.

\bibitem{Hanada:2019kue}
M.~Hanada and B.~Robinson,  \href{https://arxiv.org/abs/1911.06223}{{\ttfamily
  1911.06223}}.

\bibitem{Sundborg:1999ue}
B.~Sundborg, \href{https://doi.org/10.1016/S0550-3213(00)00044-4}{\emph{Nucl.
  Phys.} {\bfseries B573} (2000) 349}
  [\href{https://arxiv.org/abs/hep-th/9908001}{{\ttfamily hep-th/9908001}}].

\bibitem{Aharony:2003sx}
O.~Aharony, J.~Marsano, S.~Minwalla, K.~Papadodimas and M.~Van~Raamsdonk,
  \href{https://doi.org/10.4310/ATMP.2004.v8.n4.a1}{\emph{Adv. Theor. Math.
  Phys.} {\bfseries 8} (2004) 603}
  [\href{https://arxiv.org/abs/hep-th/0310285}{{\ttfamily hep-th/0310285}}].

\bibitem{Aharony:2005bq}
O.~Aharony, J.~Marsano, S.~Minwalla, K.~Papadodimas and M.~Van~Raamsdonk,
  \href{https://doi.org/10.1103/PhysRevD.71.125018}{\emph{Phys. Rev.}
  {\bfseries D71} (2005) 125018}
  [\href{https://arxiv.org/abs/hep-th/0502149}{{\ttfamily hep-th/0502149}}].

\bibitem{Hagedorn:1965st}
R.~Hagedorn, {\emph{Nuovo Cim. Suppl.} {\bfseries 3} (1965) 147}.

\bibitem{Gross:1980he}
D.~J. Gross and E.~Witten,
  \href{https://doi.org/10.1103/PhysRevD.21.446}{\emph{Phys. Rev.} {\bfseries
  D21} (1980) 446}.

\bibitem{Wadia:2012fr}
S.~R. Wadia,  \href{https://arxiv.org/abs/1212.2906}{{\ttfamily 1212.2906}}.

\bibitem{Berenstein:2002jq}
D.~E. Berenstein, J.~M. Maldacena and H.~S. Nastase,
  \href{https://doi.org/10.1088/1126-6708/2002/04/013}{\emph{JHEP} {\bfseries
  04} (2002) 013} [\href{https://arxiv.org/abs/hep-th/0202021}{{\ttfamily
  hep-th/0202021}}].

\bibitem{Bergner:2019rca}
G.~Bergner, N.~Bodendorfer, M.~Hanada, E.~Rinaldi, A.~Schafer and P.~Vranas,
  \href{https://arxiv.org/abs/1909.04592}{{\ttfamily 1909.04592}}.

\bibitem{Gupta:2007ax}
S.~Gupta, K.~Huebner and O.~Kaczmarek,
  \href{https://doi.org/10.1103/PhysRevD.77.034503}{\emph{Phys. Rev.}
  {\bfseries D77} (2008) 034503}
  [\href{https://arxiv.org/abs/0711.2251}{{\ttfamily 0711.2251}}].

\bibitem{Mykkanen:2012ri}
A.~Mykkanen, M.~Panero and K.~Rummukainen,
  \href{https://doi.org/10.1007/JHEP05(2012)069}{\emph{JHEP} {\bfseries 05}
  (2012) 069} [\href{https://arxiv.org/abs/1202.2762}{{\ttfamily 1202.2762}}].

\bibitem{Aoki:2006we}
Y.~Aoki, G.~Endrodi, Z.~Fodor, S.~D. Katz and K.~K. Szabo,
  \href{https://doi.org/10.1038/nature05120}{\emph{Nature} {\bfseries 443}
  (2006) 675} [\href{https://arxiv.org/abs/hep-lat/0611014}{{\ttfamily
  hep-lat/0611014}}].

\bibitem{Denissenya:2014poa}
M.~Denissenya, L.~{\relax Ya}. Glozman and C.~B. Lang,
  \href{https://doi.org/10.1103/PhysRevD.89.077502}{\emph{Phys. Rev.}
  {\bfseries D89} (2014) 077502}
  [\href{https://arxiv.org/abs/1402.1887}{{\ttfamily 1402.1887}}].

\bibitem{Rohrhofer:2019qwq}
C.~Rohrhofer, Y.~Aoki, G.~Cossu, H.~Fukaya, C.~Gattringer, L.~{\relax Ya}.
  Glozman et~al.,
  \href{https://doi.org/10.1103/PhysRevD.100.014502}{\emph{Phys. Rev.}
  {\bfseries D100} (2019) 014502}
  [\href{https://arxiv.org/abs/1902.03191}{{\ttfamily 1902.03191}}].

\bibitem{Alexandru:2019gdm}
A.~Alexandru and I.~Horvath,
  \href{https://arxiv.org/abs/1906.08047}{{\ttfamily 1906.08047}}.

\bibitem{Maldacena:1997re}
J.~M. Maldacena, \href{https://doi.org/10.1023/A:1026654312961,
  10.4310/ATMP.1998.v2.n2.a1}{\emph{Int. J. Theor. Phys.} {\bfseries 38} (1999)
  1113} [\href{https://arxiv.org/abs/hep-th/9711200}{{\ttfamily
  hep-th/9711200}}].

\bibitem{Witten:1998zw}
E.~Witten, \href{https://doi.org/10.4310/ATMP.1998.v2.n3.a3}{\emph{Adv. Theor.
  Math. Phys.} {\bfseries 2} (1998) 505}
  [\href{https://arxiv.org/abs/hep-th/9803131}{{\ttfamily hep-th/9803131}}].

\bibitem{Berkowitz:2016znt}
E.~Berkowitz, M.~Hanada and J.~Maltz,
  \href{https://doi.org/10.1103/PhysRevD.94.126009}{\emph{Phys. Rev.}
  {\bfseries D94} (2016) 126009}
  [\href{https://arxiv.org/abs/1602.01473}{{\ttfamily 1602.01473}}].

\bibitem{Anagnostopoulos:2007fw}
K.~N. Anagnostopoulos, M.~Hanada, J.~Nishimura and S.~Takeuchi,
  \href{https://doi.org/10.1103/PhysRevLett.100.021601}{\emph{Phys. Rev. Lett.}
  {\bfseries 100} (2008) 021601}
  [\href{https://arxiv.org/abs/0707.4454}{{\ttfamily 0707.4454}}].

\bibitem{Catterall:2008yz}
S.~Catterall and T.~Wiseman,
  \href{https://doi.org/10.1103/PhysRevD.78.041502}{\emph{Phys. Rev.}
  {\bfseries D78} (2008) 041502}
  [\href{https://arxiv.org/abs/0803.4273}{{\ttfamily 0803.4273}}].

\end{thebibliography}\endgroup

\end{document}